# High dispersive and monolithic 100% efficiency grisms

**Jörg Reinhold,[*] Marcel Schulze, Ernst-Bernhard Kley and Andreas Tünnermann**

*Institute of Applied Physics, Friedrich Schiller University Jena, Max-Wien-Platz 1, 07743 Jena, Germany*

[*]*Corresponding author: j.reinhold@uni-jena.de*

We present a type of grism, a series combination of transmission grating and prism, in which we reduce the number of diffraction orders and achieve a configuration with very high angular dispersion. The grism can be fabricated from a single dielectric material and requires no metallic or dielectric film layers for high transmission diffraction efficiency. One can reach 100% in the -1$^{st}$ transmission diffraction order and the equal damage threshold as the dielectric bulk material. We realized such an element in fused silica with an efficiency of more then 99%. The bevel backside reflection is reduced by a statistical antireflective structure, so we measured an efficiency of the entire grism of 95% at a single wavelength.

*OCIS codes:* 050.1940, 050.1950, 050.1970, 050.6624

Diffraction gratings are an integral part of many modern optical systems with manifold applications in laser systems, spectroscopy and telecommunication. For example in laser applications they provide wavelength selectivity for tunable lasers, pulse compression for ultrafast lasers and power scaling for wavelength-combined lasers. Performance goals of those diffraction gratings are for example dispersion, efficiency, damage threshold and bandwidth.

A diffraction grating is a periodically varying boundary between two materials with different refractive indices $n_1$ and $n_2$. The correlation between incident angle $\Theta_{in}$ and transmission diffraction angle $\Theta_m$ of $m^{th}$ order is given by the grating equation

$$n_1 \sin(\Theta_{in}) - n_2 \sin(\Theta_m) + \frac{m\lambda}{\Lambda} = 0, \qquad (1)$$

wherein $\lambda$ is the wavelength and $\Lambda$ is the grating period. Angles are measured with respect to the grating normal. As the spectral properties of the grating are determined by Eq. (1), the efficiency of the grating depends on the profile of the periodic structure and can be predicted by rigorous calculations, e.g. by rigorous coupled wave algorithm (RCWA) [1]. For binary gratings the structure is determined by the parameters duty cycle and groove depth.

The theoretical damage threshold of an Au grating is 1.5 J/cm² for ns-pulses and 0.5 J/cm² for ps-pulses (at $\lambda$ = 700 nm to 1,100 nm) [2]. For a dielectric grating the damage threshold raises to 20 J/cm² for ns-pulses and 2 J/cm² for 400 fs pulses (at $\lambda$ = 1,053 nm) [3]. Therefore dielectric gratings are replacing commercially available metal-coated reflection gratings for high power short pulse laser systems.

Furthermore dielectric gratings have the capability to very high diffraction efficiencies. Commonly, to achieve a high efficiency, transmission gratings will be used in Littrow configuration where the -1$^{st}$ reflection order will propagate antiparallel to the incident wave. In calculations 98% efficiency is possible [4], while in laboratory 95% are measured with a fused silica grating [5].

This high efficiency decrease significantly when high dispersion is required. The grating equation (Eq. 1) results in

$$D = \left|\frac{d\Theta_m}{d\lambda}\right| = \left|\frac{m}{\Lambda \cos(\Theta_m)}\right| \qquad (2)$$

for the angular dispersion *D*. Thus a higher dispersion is possible by decreasing the period. A smaller period induces a larger angle of incidence for fulfilling the Littrow configuration. This leads to larger reflection losses. As an example, at a wavelength of 1064 nm, the profile of a rectangular fused silica transmission grating (refractive index 1.45) with D = 0.06 °/nm can be optimized theoretically to diffract more than 98% to the -1$^{st}$ transmitted order. On contrary, a grating with D = 0.12 °/nm cannot exhibit more than 93%.

By the use of dielectric film layers this efficiency can be increased. Efficiencies up to 96% for high dispersive gratings where measured [6]. The different materials in the element reduce the damage threshold introduced by lasers or temperature fluctuations. However, there is an 'one material concept' for a high dispersion and high efficiency transmission grating published [7]. This grating is embedded in fused silica, which allows for an efficient suppression of any reflection losses. To embed the grating, a plane fused silica plate was bonded on the grating. A theoretical efficiency of 100% were predicted and 98% were measured with a grating in fused silica (n = 1.45) at a wavelength of 1064 nm, while an angle dispersion of 0.12 °/nm were determined.

Here we present another concept. It shows high efficiencies combined with high angular dispersion. With this concept there is no need for a bonding step like in the encapsulated concept. The sample can be fabricated from a single dielectric bulk material. As opposed to the above named concepts we do not use the Littrow configuration. Here only three propagating orders exist: the 0$^{th}$ order in reflection and the 0$^{th}$ and -1$^{st}$ order in transmission. There is no -1$^{st}$ reflection order like in Littrow configuration. Accordingly the -1$^{st}$ diffraction order does not transmit through the substrate's backside cause of total internal reflection (TIR).

Possess the element a bevel backside, then the beam can propagate below the angle of TIR

$$\Theta_{TIR} = \arcsin\left(\frac{n_1}{n_2}\right)$$

and light can transmit. Here $n_1$ is the diffractive index of the surrounding medium and $n_2$ the diffractive index of the substrate. See Fig. 1.a for the principle beam propagations of major passes inside and outside the substrate.

Diffraction efficiencies in dependence of duty cycles and groove depths were calculated with RCWA for TM-polarization. For comparison three grating concepts were calculated with the same grating period of 0.51λ and the same dispersion at each wavelength. For example at 1064 nm the dispersion D is 0.15 °/nm and at 671 nm (our experimental wavelength) D is 0.24 °/nm. The used refractive index of the substrate's material is that for fused silica, $n_2 = 1.45$. In contrast to the conventional surface relief grating (see Fig. 2.a), the diffraction efficiency of our grating concept predicts a broad maximum higher than 95% (see Fig. 2.c). Like for the encapsulated grating (see Fig. 2.b) it achieves 100% for particular parameters. These optimized parameters are a duty cycle of 0.42 and a grove depth of 0.74λ at an incident angle of 68.82°. The -1$^{st}$ diffraction order has a diffraction angle of 45° in the substrate.

The grating can also be used in the opposite direction. So the incident wave firstly propagates through the substrate and is then diffracted by the grating (see Fig.1.b). The diffraction efficiency for the -1$^{st}$ order will be also 100% for optimized parameters.

As yet the Fresnel reflections were unmentioned. In opposite to the encapsulated grating here we only have one reflective side of the sample, the backside. One idea to reduce the reflection is a dielectric layer system. Unfortunately, this leads to a lower damage threshold than the bulk dielectric, but still has a higher threshold than a metallic layer system. To reach the same damage threshold as the bulk material, we use statistical antireflective nanostructures [8, 9].

The grism was realized in fused silica. For measurements we use a cw-laser at 671 nm. So from design considerations the grating period should be 342 nm and the groove depth should be 496nm. We used electron beam lithography for grating pattern generation, which provides highly flexibility for parameter variations, especially period and duty cycle. In fabrication a 25 mm x 25 mm fused silica substrate was coated firstly by a chromium layer and then by an electron resist. This sample was exposed to the electron beam. After the electron beam writing the resist was developed. The resulting resist grating was etched into the chromium layer by reactive ion etching (RIE). This chromium pattern acts as a hard mask for the pattern transfer into a fused silica substrate by reactive ion beam etching (RIBE). Then the bevel was realized by polishing the backside. So we achieve an angle of $(9 \pm 1)$ ° with respect to the grating surface.

For measurements the incoming wave propagates firstly through the substrate like in Fig. 1.b. We measured the Fresnel reflection of the incident wave and the -1$^{st}$ transmitted order at 671 nm and TM-polarization. We measured in the -1$^{st}$ transmission diffraction order $(85 \pm 1)$ % and a Fresnel reflections of $(14 \pm 1)$ % so the grating efficiency is $(99 \pm 1)$ %.

After these measurements we realized the antireflective structures at the bevel backside of the grism. Therefore we used a special dry etching technique to establish capable antireflective nanostructures. In this way the surface achieves a continuous refractive index transition between the substrate and air. The structures are smaller than the smallest and deeper than the half of the largest occurring wavelengths. For the assembly of these statistical structures we neither need any lithographic methods, like electron beam lithography or photolithography, nor any other mask fabrication. Trough the antireflective structure we can reduced the measured Fresnel reflection to $(5 \pm 1)$ %. The entire grism has now a measured efficiency of $(94 \pm 1)$ %.

Results of the spectral and angular measurements of the sample are illustrated in Fig. 5 and Fig. 6. The measured efficiencies represent an average over a measurement spot of about 1 mm x 1 mm on the sample. For comparison, the gray curves illustrate theoretical simulations of a sample with a perfect antireflective backside and the dashed grey curves shows a conventional surface relief structure with the same angular dispersion.

We have shown a concept to reach a diffraction element which is also high efficient and high dispersive. Theoretically it can diffract 100% in the $-1^{st}$ diffraction order with at a single wavelength and a angular dispersion of 15 °/mm. The element can be fabricated from a single dielectric substrate. That's why the element has a high damage threshold. We have fabricated an element on with a grating efficiency of $(99 \pm 1)$ %. Through a statistically antireflective strucure we have reduced the backside reflection we have measured an efficiency of $(95 \pm 1)$ of the entire grism structure.

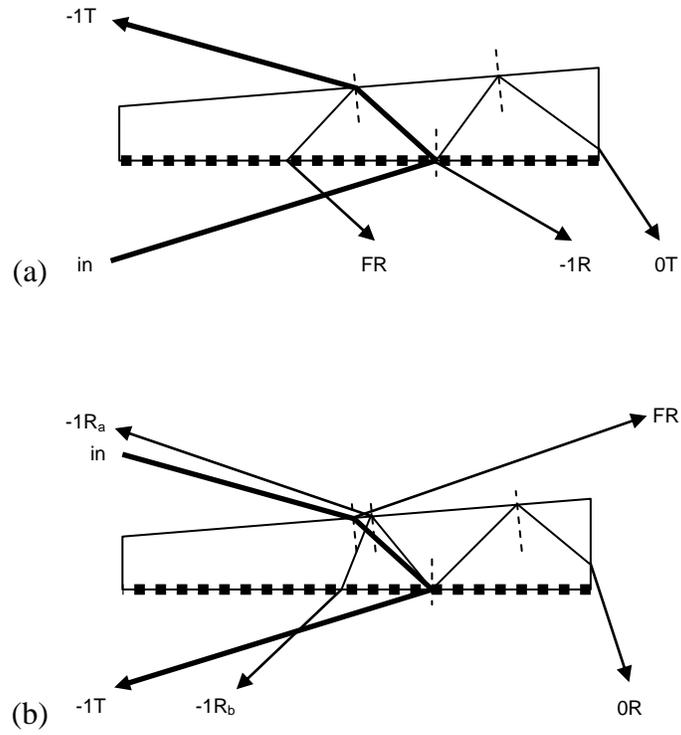

Fig. 1. Beam propagation of major passes. (a) First the beam passes the grating and then the substrate. (b) First the beam passes the substrate and then the grating.

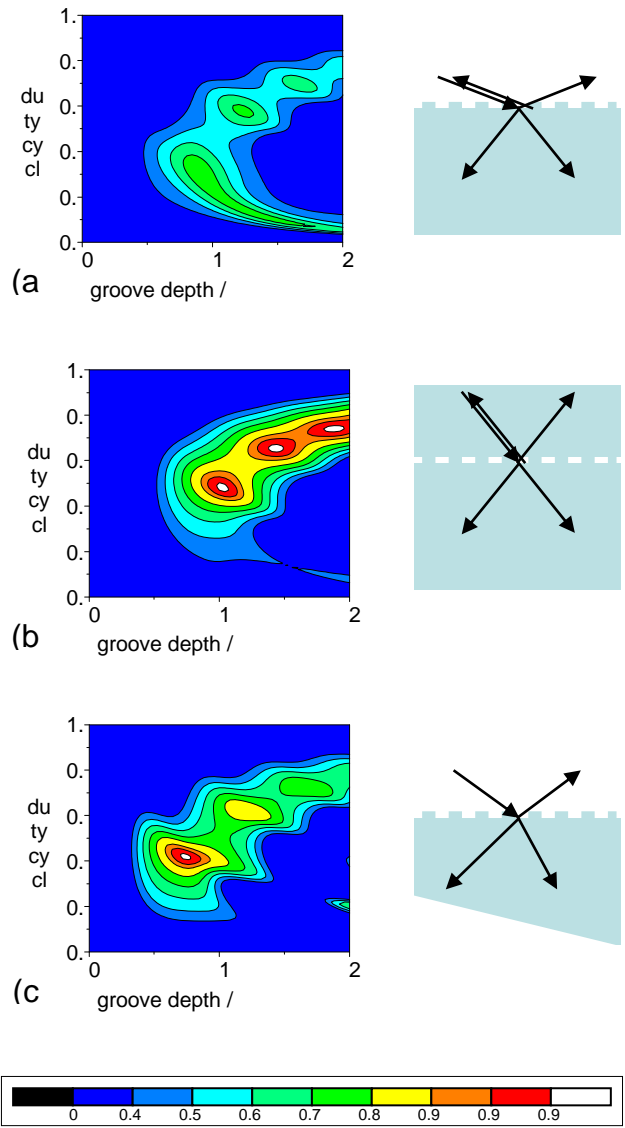

Fig. 2. Efficiencies of 1$^{st}$ transmission diffraction order for different concepts dependence on duty cycles and groove depths. All have the same angular dispersion. The first concept (a) is a simple rectangular surface grating used in Littrow configuration. The second concept (b) works with an encapsulated grating used in Littrow configuration. The third concept (c) is the grism concept.

9